%% file: paper.tex
\documentclass[aps,prd,final,groupedaddress,showpacs,showkeys,floatfix,preprintnumbers,nofootinbib,letterpaper,notitlepage]{revtex4-1}

\usepackage{graphicx}
\usepackage{float}

\usepackage[latin1]{inputenc}                    
\usepackage{latexsym}                            
\usepackage{amsfonts}                            
\usepackage{amssymb}                             
\usepackage{amsmath}                             
\usepackage[mathscr]{eucal}                      
\usepackage{dcolumn}                             
\usepackage{subfigure}
\usepackage[english]{babel}
\usepackage{theorem}                            
\usepackage{hyperref}
\usepackage{slashed}
\usepackage{color}
\usepackage{dcolumn}

\input{defs}

\begin{document}

\color{black}

\input{titlepage}

\section{\label{sect:Introduction}Introduction}

\input{introduction}

\section{\label{sect:lqcd_method}Lattice methodology}

\input{methodology}

\section{Results}

\input{results}

\section{Summary and conclusions}

\input{conclusions}

\begin{acknowledgments}

We thank Zoltan Fodor for useful discussions and the
Budapest-Marseille-Wuppertal collaboration for making some of their
configurations available to us.  This research used resources of the
Argonne Leadership Computing Facility at Argonne National Laboratory,
which is supported by the Office of Science of the U.S. Department of
Energy under contract \#DE--AC02--06CH11357, resources provided by
the New Mexico Computing Applications Center (NMCAC) on Encanto,
resources at Forschungszentrum J\"ulich, and facilities of the USQCD
Collaboration, which are funded by the Office of Science of the
U.S.~Department of Energy.

During this research JRG, JWN and AVP were supported in part by the
U.S.~Department of Energy Office of Nuclear Physics under grant
\#DE--FG02--94ER40818, ME was supported in part by DOE grant
\#DE--FG02--96ER40965, SNS was supported in part by DOE contract
\#DE--AC02--05CH11231, and SK was supported in part by Deutsche
Forschungsgemeinschaft through grant SFB--TR~55.

The Chroma software suite \cite{Edwards:2004sx} was used for the mixed action and
DWF calculations. The clover Wilson calculations were performed with Qlua software.

\end{acknowledgments}

\appendix
\section{\label{sect:app-npr}
  Non-perturbative renormalization of the tensor charge operator}

\input{app-npr}

\section{\label{sect:app-chipt}Chiral Perturbation Theory}
\input{chipt_scalar}

\input{chipt_tensor}

\section{Table of results}

\input{table}

\bibliography{paper}

\printtables

\printfigures

\end{document}

%% file: defs.tex
\newcommand{\beq}{\begin{equation}}
\newcommand{\eeq}{\end{equation}}
\newcommand{\bgath}{\begin{gathered}}
\newcommand{\egath}{\end{gathered}}
\newcommand{\bsplit}{\begin{split}}
\newcommand{\esplit}{\end{split}}
\newcommand{\bdm}{\begin{displaymath}}
\newcommand{\edm}{\end{displaymath}}
\newcommand{\beqn}{\begin{eqnarray}}
\newcommand{\eeqn}{\end{eqnarray}}
\newcommand{\bea}[1]{\beq\begin{array}{#1}}
\newcommand{\eea}{\end{array}\eeq}
\newcommand{\bi}{\begin{itemize}}
\newcommand{\ei}{\end{itemize}}
\newcommand{\ben}{\begin{enumerate}}
\newcommand{\een}{\end{enumerate}}
\newcommand{\ba}[1]{\begin{array}{#1}}
\newcommand{\ea}{\end{array}}

\newcommand{\bc}{\begin{center}}
\newcommand{\ec}{\end{center}}
\newcommand{\bfr}{\begin{flushright}}
\newcommand{\efr}{\end{flushright}}
\newcommand{\bfl}{\begin{flushleft}}
\newcommand{\efl}{\end{flushleft}}

\newcommand{\bv}{\begin{verbatim}}
\newcommand{\ev}{\end{verbatim}}

%% file: titlepage.tex
\preprint{MIT-CTP 4359}
\preprint{WUB/12-07}
\preprint{UCB-NPAT-12-008}
\preprint{NT-LBL-12-008}
\title{Nucleon Scalar and Tensor Charges from Lattice QCD with Light Wilson Quarks}

\author{J.~R.~Green}
\author{J.~W.~Negele}
\author{A.~V.~Pochinsky}
  \affiliation{Center for Theoretical Physics, Massachusetts Institute of Technology, 
               Cambridge, Massachusetts 02139, USA}
\author{ S.~N.~Syritsyn} 
  \affiliation{Nuclear Science Division, Lawrence Berkeley National Laboratory, 
               Berkeley, California 94720, USA}

\author{M.~Engelhardt}
  \affiliation{Department of Physics, New Mexico State University, Las
    Cruces, New Mexico 88003, USA}

\author{S.~Krieg}
  \affiliation{Bergische Universit\"at Wuppertal, D-42119 Wuppertal, Germany}
  \affiliation{IAS, J\"ulich Supercomputing Centre, Forschungszentrum
    J\"ulich, D-52425 J\"ulich, Germany}
   
   \date{\today}

\begin{abstract}

  We present 2+1~flavor Lattice QCD calculations of the nucleon scalar
  and tensor charges. Using the BMW clover-improved Wilson action with
  pion masses between 149 and 356~MeV and three source-sink
  separations between 0.9 and 1.4~fm, we achieve good control over
  excited-state contamination and extrapolation to the physical pion
  mass. As a consistency check, we also present results from
  calculations using unitary domain wall fermions with pion masses
  between 297 and 403~MeV, and using domain wall valence quarks and
  staggered sea quarks with pion masses between 293 and 597~MeV.

\end{abstract}

\pacs{
12.38.Gc, 
12.60.-i, 
13.30.Ce, 
14.20.Dh}

\maketitle

%% file: introduction.tex
A quantitative understanding of the quark structure of hadrons is
essential to relate experimentally observed hadronic properties to
fundamental processes occurring at the quark level.  Thus, Lattice QCD
can play a crucial role in determining parameters of the Standard
Model from experiment and in predicting the experimental effects of
interactions beyond the Standard Model.
 
One example is the search for the effects of new scalar and tensor
couplings beyond the familiar weak interactions of the Standard Model
in the decay of ultra-cold neutrons as discussed in detail in
\cite{Bhattacharya:2011qm}.  These couplings would contribute through
the matrix elements $\langle p | \bar{u} d | n \rangle $ and $\langle
p | \bar{u} \sigma_{\mu \nu } d | n \rangle $.  To leading order in
the recoil approximation, these matrix elements are proportional to
the isovector scalar and tensor charges $g_S $ and $g_T $ of the
nucleon.  Hence, Lattice QCD predictions of these charges are crucial
for determining the sensitivity of the decay process to the
non-standard couplings.

Another example concerns the search for an electric dipole moment of
the neutron, which constrains the possible values of the CP-violating
strong interaction $\theta $ angle.  The constant of proportionality
between the electric dipole moment and $\theta $ can be calculated
directly using a background electric field
\cite{Shintani:2006xr,Shintani:2008nt} or from the CP-odd form factor
$F_3 $, cf.~\cite{Aoki:2008gv,Liu:2008gr}.  However, it is also
instructive to relate the neutron electric dipole moment near the
chiral limit at small $\theta $ to the CP-violating pion-nucleon
coupling constant $\bar{g}_{\pi NN} $,
cf.~\cite{Crewther:1979pi}. This assumes that $|\pi N \rangle $
intermediate states provide the dominant contribution to the dipole
moment, which is the case as $m_{\pi } \rightarrow 0$. In turn,
$\bar{g}_{\pi NN} $ is proportional to the isovector scalar
charge $g_S $ of the nucleon \cite{Crewther:1979pi}, which thus again
appears as an important parameter influencing the search for new
physics. Given that the physical limit occurs at finite $m_{\pi } $,
where there will be corrections to the relation between the electric
dipole moment of the neutron and $\bar{g}_{\pi NN} $, a comparison of
this approximation to direct evaluations can also provide insight into
the importance of these corrections.  Another important effect of
$\bar{g}_{\pi NN} $ is the fact that it induces CP violation in
nuclear forces resulting in strongly enhanced nuclear electric dipole
moments in certain cases \cite{Haxton:1983dq}.

These two examples directly motivate the present work in which we
calculate the nucleon's isovector charges $g_S $ and $g_T $. Since the
computationally formidable disconnected contributions from couplings
to the sea quarks cancel in these isovector charges, we only calculate
the connected contributions.  Extensive Lattice QCD data generated
with three different lattice actions at pion masses ranging between
149 MeV and 597 MeV are reported, together with their extrapolations
to the physical pion mass.

In addition to the primary motivation described above, our results
also contribute to a more detailed understanding of the nucleon sigma
term $\sigma_{N} =\frac{1}{2} (m_u + m_d ) \langle N| \bar{u} u
+\bar{d} d |N\rangle $, which determines the part of the nucleon mass
generated by the light quark degrees of freedom via the spontaneous
breaking of chiral symmetry. Whereas we do not have the resources to
evaluate the disconnected contributions to $\sigma_{N} $ at present,
the connected contributions to $\langle N| \bar{u} u |N\rangle $ and
$\langle N| \bar{d} d |N\rangle $ can be given separately and are
expected to furnish most of the strength of the nucleon sigma term;
cf.~\cite{Ohki:2008ff,Durr:2011mp,Horsley:2011wr} for recent lattice
studies focusing on $\sigma_{N} $. These quantities are also relevant
to dark matter searches based on Higgs-mediated couplings of baryonic
to dark matter \cite{Bottino:2001dj,Ellis:2005mb,Bertone:2010zz}. For
example, the neutralino-nucleon scalar cross-section considered in
\cite{Bottino:2001dj} can be cast as a linear combination of terms
proportional to $\langle N | \bar{u} u | N \rangle $ and $\langle N |
\bar{d} d | N \rangle $. Thus, again, the scalar charge $g_S $ of the
nucleon (including both isoscalar and isovector components) is useful
in quantifying searches for these dark matter candidates.

%% file: methodology.tex
The full set of lattice ensembles on which we have calculated our
observables is listed in Tab.~\ref{tab:ensembles}. We make use of
three different lattice actions.

\begin{table}
  \centering
  \begin{tabular}{clcD{?}{}{5.4}crr}
    Label & $a$ [fm] & $L_x^3\times L_t$ & \multicolumn{1}{c}{$m_\pi$ [MeV]} & $m_{\pi } L_x a$ & \# confs & \# meas \\\hline
    \multicolumn{7}{c}{Wilson-clover} \\
    W1 & 0.09  & $32^3\times 64$ & 317?(2) & 4.6 & 103 & 824 \\
    W2 & 0.116 & $48^3\times 48$ & 149?(1) & 4.2 & 646 & 7752 \\
    W3 & 0.116 & $32^3\times 48$ & 202?(1) & 3.8 & 457 & 5484 \\
    W4 & 0.116 & $32^3\times 96$ & 253?(1) & 4.8 & 202 & 2424 \\
    W5 & 0.116 & $32^3\times 48$ & 254?(1) & 4.8 & 420 & 5040 \\
    W6 & 0.116 & $24^3\times 48$ & 254?(1) & 3.6 & 418 & 10032 \\
    W7 & 0.116 & $24^3\times 48$ & 303?(2) & 4.3 & 128 & 768 \\
    W8 & 0.116 & $24^3\times 48$ & 356?(2) & 5.0 & 127 & 762 \\\hline
    \multicolumn{7}{c}{Domain wall} \\
    D1 & 0.084 & $32^3\times 64$ & 297?(5) & 4.0 & 615 & 4920 \\
    D2 & 0.084 & $32^3\times 64$ & 355?(6) & 4.8 & 882 & 7056 \\
    D3 & 0.084 & $32^3\times 64$ & 403?(7) & 5.5 & 527 & 4216 \\
    D4 & 0.114 & $24^3\times 64$ & 329?(5) & 4.6 & 399 & 3192 \\\hline
    \multicolumn{7}{c}{Mixed action} \\
    M1 & 0.124 & $20^3\times 64$ & 293?(6) & 3.7 & 460 & 3680 \\
    M2 & 0.124 & $28^3\times 64$ & 356?(7) & 6.3 & 272 & 2176 \\
    M3 & 0.124 & $20^3\times 64$ & 356?(7) & 4.5 & 628 & 5024 \\
    M4 & 0.124 & $20^3\times 64$ & 495?(10)& 6.2 & 483 & 3864 \\
    M5 & 0.124 & $20^3\times 64$ & 597?(12)& 7.5 & 562 & 4496 \\
  \end{tabular}
  \caption{Lattice ensembles used for scalar and tensor charge calculations.}
  \label{tab:ensembles}
\end{table}

Our main results are calculated on ensembles at light pion masses,
with tree-level clover-improved Wilson fermions coupled to double
HEX-smeared gauge fields, as used by the BMW collaboration
\cite{Durr:2010aw}. Operators are renormalized nonperturbatively using
the Rome-Southampton method. For the scalar charge, we use
renormalization factors calculated by the BMW collaboration
\cite{Durr:2010aw}, and for the tensor charge we performed our own
calculation, see Appendix~\ref{sect:app-npr}. 
These factors are given in Tab.~\ref{tab:BMW_renorm}.

\begin{table}[htb]
  \centering
  \begin{tabular}{lD{.}{.}{4.10}D{.}{.}{4.8}}
    $a$ [fm] & \multicolumn{1}{c}{$Z_S$} & \multicolumn{1}{c}{$Z_T$} \\\hline
    0.09 & 1.107(16)(22) & 1.011(5)\\
    0.116& 1.115(17)(30) & 0.9624(62)
  \end{tabular}
  \caption{$\overline{MS}$ renormalization factors at $\mu=2$~GeV for
    the tensor and scalar charge on the Wilson-clover ensembles.}
\label{tab:BMW_renorm}
\end{table}

For comparison and for a consistency check, we also present results on
ensembles with two different lattice actions used for earlier nucleon
structure calculations. First, we use unitary domain wall quarks on
ensembles generated by the RBC and UKQCD collaborations
\cite{Allton:2008pn,Aoki:2010dy}. Details of our analysis methods are
in Ref.~\cite{Syritsyn:2009mx}. Operators are renormalized
nonperturbatively using the Rome-Southampton method. $Z_S$
\cite{Aoki:2010dy} and, on the coarse ensemble, $Z_T$
\cite{Aoki:2010xg}, were calculated by the RBC collaboration. On the
fine ensembles, we found $Z_T^{\overline{MS}(2\text{
    GeV})}=0.8168(9)$, see Appendix~\ref{sect:app-npr}.

Second, our mixed-action scheme \cite{Bratt:2010jn,Hagler:2007xi} uses domain
wall valence quarks on gauge configurations with Asqtad staggered sea
quarks generated by the MILC collaboration \cite{Bernard:2001av}.
Renormalization factors are calculated in the same
way~\cite{Bistrovic_thesis} as those used in previous calculations of
nucleon generalized form factors.  One-loop perturbation theory is
used to calculate the renormalization factors for all quark bilinear
operators evaluated at the one-loop coupling constant and $\mu^2 =
1/a^2$.  Because HYP smearing suppresses loop integrals, the ratio $
Z_{\mathcal{O},\text{pert}}/ Z_{A,\text{pert}}$, where $\mathcal{O}$ denotes a
general bilinear and $A$ is the axial current, is within a few percent
of unity indicating perturbative corrections are already small at the
one-loop level for this ratio. The multiplicative wave function
renormalization in the fifth dimension appearing in all
renormalization factors is included non-perturbatively by using the
five-dimensional conserved axial current for domain wall fermions to
calculate $Z_{A,\text{nonpert}}$ and calculating $Z_{\mathcal{O}}=(
Z_{\mathcal{O},\text{pert}} /Z_{A,\text{pert}}) \cdot Z_{A,\text{nonpert}}$.  The relevant
matrix elements are~\cite{Bistrovic_thesis} $Z_{A,\text{pert}} = 0.964$,
$Z_{S,\text{pert}} = 0.971$, and $Z_{T,\text{pert}} = 0.987$, and the result is
evolved to $\mu = 2$~GeV.

For the scalar charge, we can alternatively make use of the
renormalization group invariant combination
$(m_s-m_{ud})g_S$. Dividing by the physical $m_s-m_{ud}$, as measured
in 2+1~flavor Lattice QCD calculations in the $\overline{MS}$ scheme
at 2~GeV~\cite{Colangelo:2010et,Bazavov:2009fk,Bazavov:2010yq,McNeile:2010ji,Aoki:2010dy,Durr:2010vn}, and multiplying by a ratio of
differences of pseudoscalar meson masses to cancel the leading
dependence on quark masses, we get
\begin{equation}
  g_S^{\overline{MS}(2\text{ GeV})} \approx \frac{(m_s^\text{bare}-m_{ud}^\text{bare})g_S^\text{bare}}{m_s^\text{phys}-m_{ud}^\text{phys}}\times\frac{m_{K,\text{phys}}^2-m_{\pi,\text{phys}}^2}{m_K^2-m_\pi^2},
\end{equation}
where the physical $m_K$ and $m_\pi$ in the isospin limit are from
Ref.~\cite{Colangelo:2010et} and their values on our mixed-action
ensembles were computed in an earlier
work~\cite{WalkerLoud:2008bp}. On the mixed-action ensembles, this
approach yields values of $g_S$ that differ from the perturbatively
renormalized values by between 0.4\% and 1.3\%. Assuming the
perturbative renormalization of the tensor charge has similar errors,
we use our perturbative approach for both $g_S$ and $g_T$ and
conservatively estimate a systematic error due to renormalization of
2\%.

We compute nucleon forward matrix elements using the usual
ratio-plateau method \cite{Bratt:2010jn,Syritsyn:2009mx}. Beginning
with two-point and three-point functions,
\begin{align}
C_\text{2pt}(t,\vec P) &= \langle N(\vec p=\vec P, t) \bar N(\vec x=0, 0)\rangle, \\
C_\text{3pt}^{\mathcal{O}}(\tau,T;\vec P) &= \langle N(\vec p=\vec P, T) \mathcal{O}(\vec p=0, \tau) \bar N(\vec x=0, 0)\rangle,
\end{align}
where $N$ is our lattice nucleon interpolating operator, we
compute their ratio,
\begin{equation}
R^\mathcal{O}(\tau,T;\vec P) = \frac{C_\text{3pt}^\mathcal{O}(\tau,T;\vec P)}{C_\text{2pt}(T,\vec P)}.
\end{equation}
At sufficiently large $\tau$ and $T-\tau$, contributions from excited
states are negligible, and the ratio gives us the matrix element
$\langle N(\vec P)|\mathcal{O}|N(\vec P)\rangle$.  In practice, for a
fixed source-sink separation $T$, we take the average over the central
two or three points of the plateau as the matrix element. The matrix
elements give us the isovector scalar and tensor charges:
\begin{align}
  \langle N(\vec P)|\bar u u - \bar d d|N(\vec P)\rangle &= g_S \bar u(\vec P)u(\vec P) \\
  \langle N(\vec P)|\bar u \sigma_{\mu\nu} u - \bar d \sigma_{\mu\nu} d|N(\vec P)\rangle &= g_T \bar u(\vec P)\sigma_{\mu\nu}u(\vec P).
\end{align}
We take the weighted average of results measured with $\vec P=0$ and
$\vec P = \tfrac{2\pi}{L}(-1,0,0)$. The source-sink separations $T$
that we use are listed in Tab.~\ref{tab:sourcesink}. On the Wilson
action ensembles, we perform measurements at three source-sink
separations in order to better identify systematic errors from
excited-state contamination. However, our main results are presented
using the middle separation, with $T\approx 1.16$~fm.

\begin{table}[htb]
  \centering
  \begin{tabular}{llc}
    Action & $a$ [fm] & $T/a$\\\hline
    Mixed & 0.124 & 9 \\
    Domain wall & 0.084 & 12 \\
    Domain wall & 0.114 & 9 \\
    Wilson-clover & 0.09 & 10, 13, 16 \\
    Wilson-clover & 0.116 & 8, 10, 12
  \end{tabular}
  \caption{Source-sink separations $T$.}
  \label{tab:sourcesink}
\end{table}

Chiral perturbation theory results for $g_A$ and $g_T$ are summarized
in Appendix~\ref{sect:app-chipt}.

%% file: results.tex
\begin{figure}
  \centering
  \includegraphics{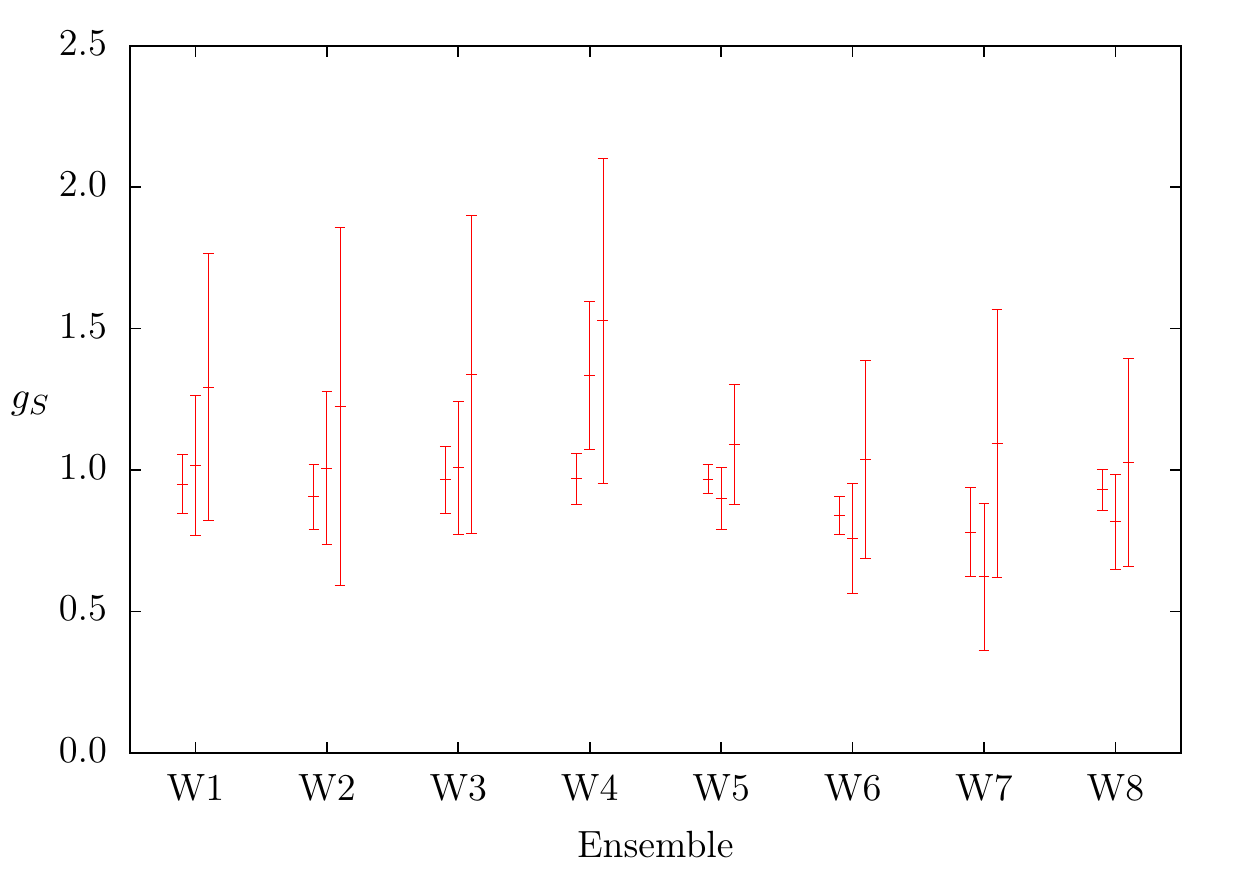}
  \caption{Scalar charge measured at three source-sink separations on the
    Wilson-clover ensembles, as enumerated in Tab.~\ref{tab:ensembles}.}
\label{fig:gSsep}
\end{figure}
\begin{figure}
  \centering
  \includegraphics{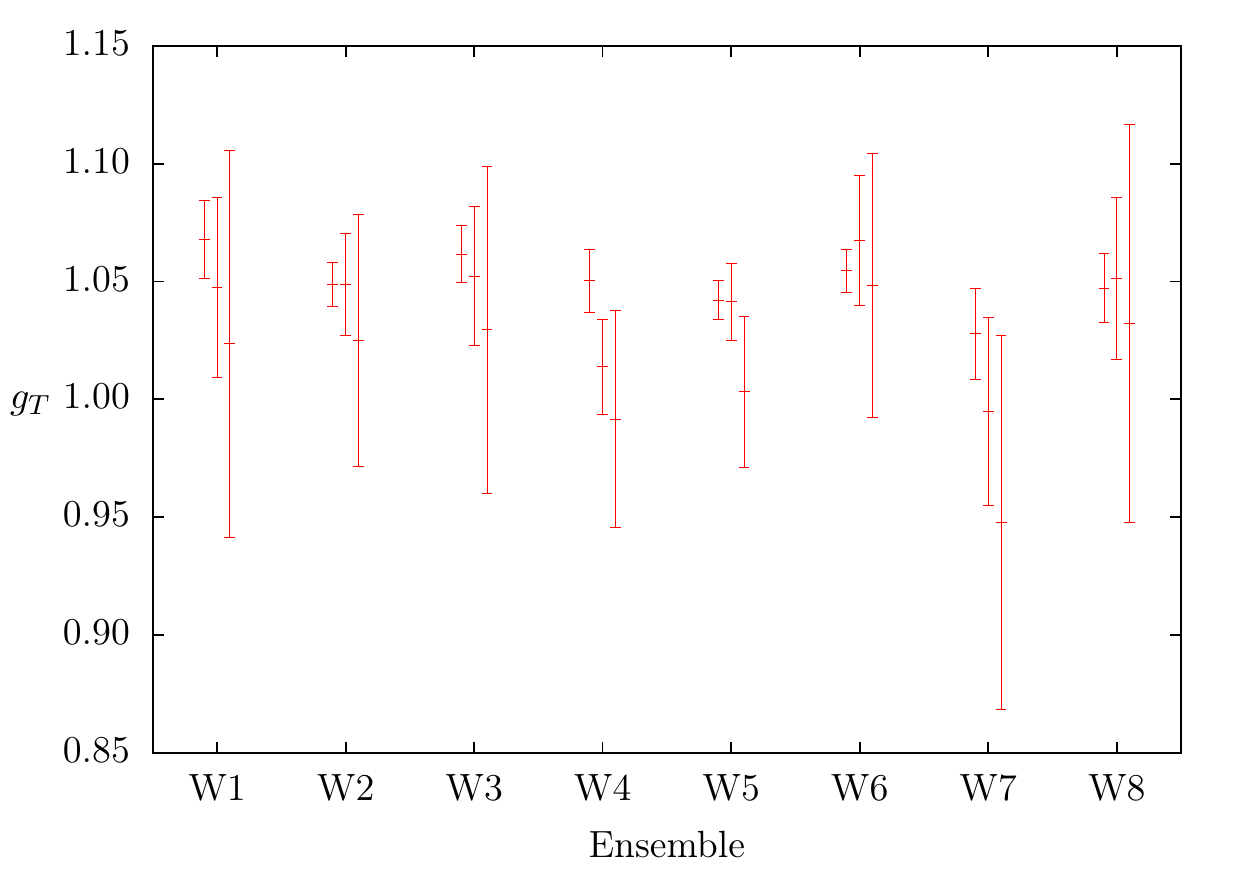}
  \caption{Tensor charge measured at three source-sink separations on the
    Wilson-clover ensembles, as enumerated in Tab.~\ref{tab:ensembles}.}
\label{fig:gTsep}
\end{figure}

The dependence on source-sink separation for the
Wilson-clover data is shown in Fig.~\ref{fig:gSsep} for the
scalar charge and in Fig.~\ref{fig:gTsep} for the tensor charge.
Neither observable shows a strong statistically significant dependence
on source-sink separation.

\begin{figure}
  \centering
  \includegraphics{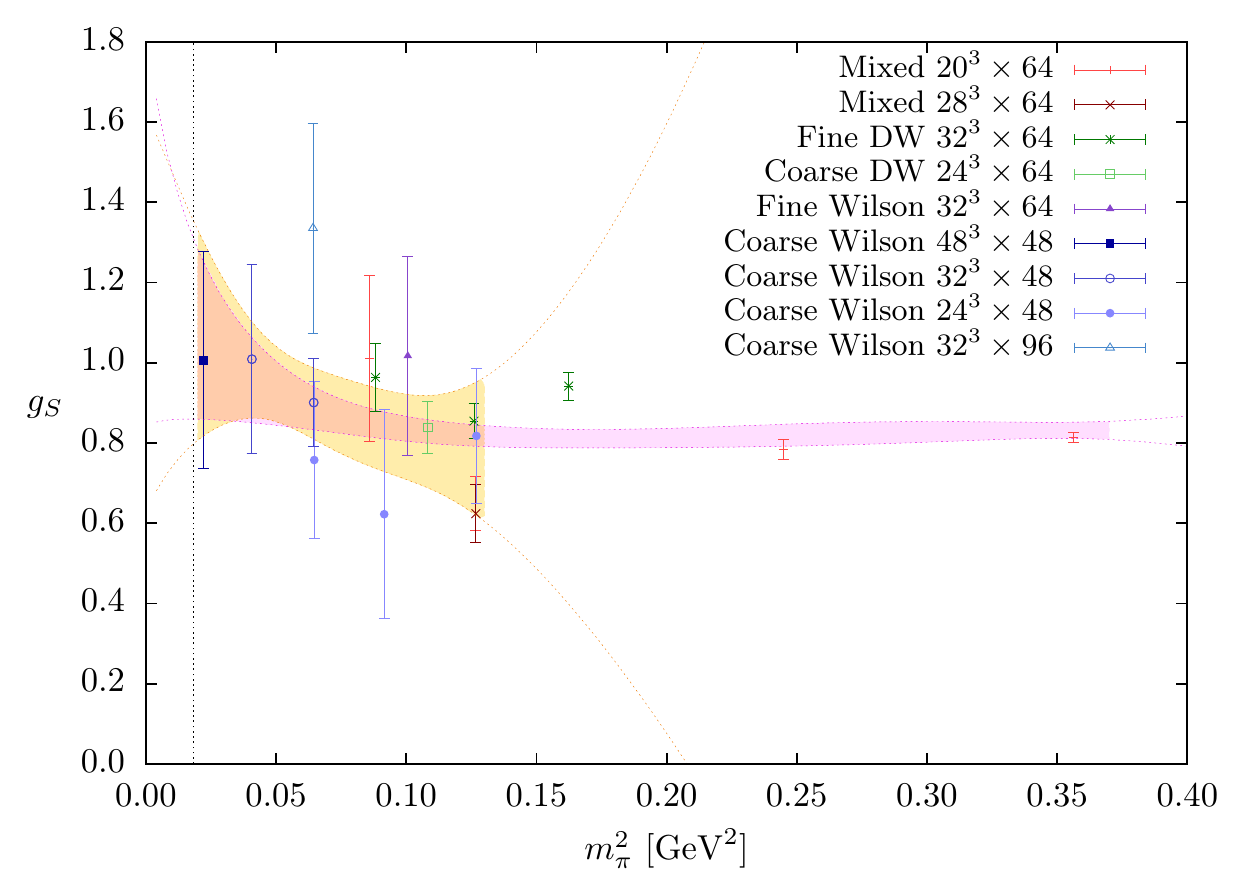}
  \caption{$g_S$ versus $m_\pi^2$. Error bars are purely statistical
    and do not include errors in renormalization factors, which are
    correlated across ensembles with the same action. Two chiral fits
    are shown: a three-parameter fit to the coarse Wilson-clover
    ensembles (orange) and a four-parameter fit to all shown ensembles
    (magenta).}
\label{fig:gSmpi}
\end{figure}

The main results for the scalar charge are presented in
Fig.~\ref{fig:gSmpi}. There is broad agreement between the different
actions, although they are not entirely consistent. In particular, the
two mixed-action ensembles at $m_\pi=356$~MeV have points that lie
significantly below the point for the domain wall ensemble at
$m_\pi=355$~MeV. We estimate the value of the scalar charge at the
physical pion mass, 135~MeV, in three ways:
\begin{enumerate}
\item Taking the value from our ensemble with the lowest pion mass,
  149~MeV. Including the error in $Z_S$, this is $g_S=1.01(27)$.
\item Performing a 3-parameter chiral fit to the set of seven
  $a=0.116$ Wilson-clover ensembles. This is a good fit with
  $\chi^2/\text{dof}=3.99/4$ and is shown in
  Fig.~\ref{fig:gSmpi}. Extrapolating to the physical pion mass yields
  $g_S=1.08(28)$.
\item Performing a 4-parameter chiral fit to the full set of
  ensembles. The tension at higher pion masses between the domain wall
  and the mixed action ensembles is reflected in
  $\chi^2/\text{dof}=30.23/13$. Extrapolation yields $g_S=1.08(23)$.
\end{enumerate}

\begin{figure}
  \centering
  \includegraphics{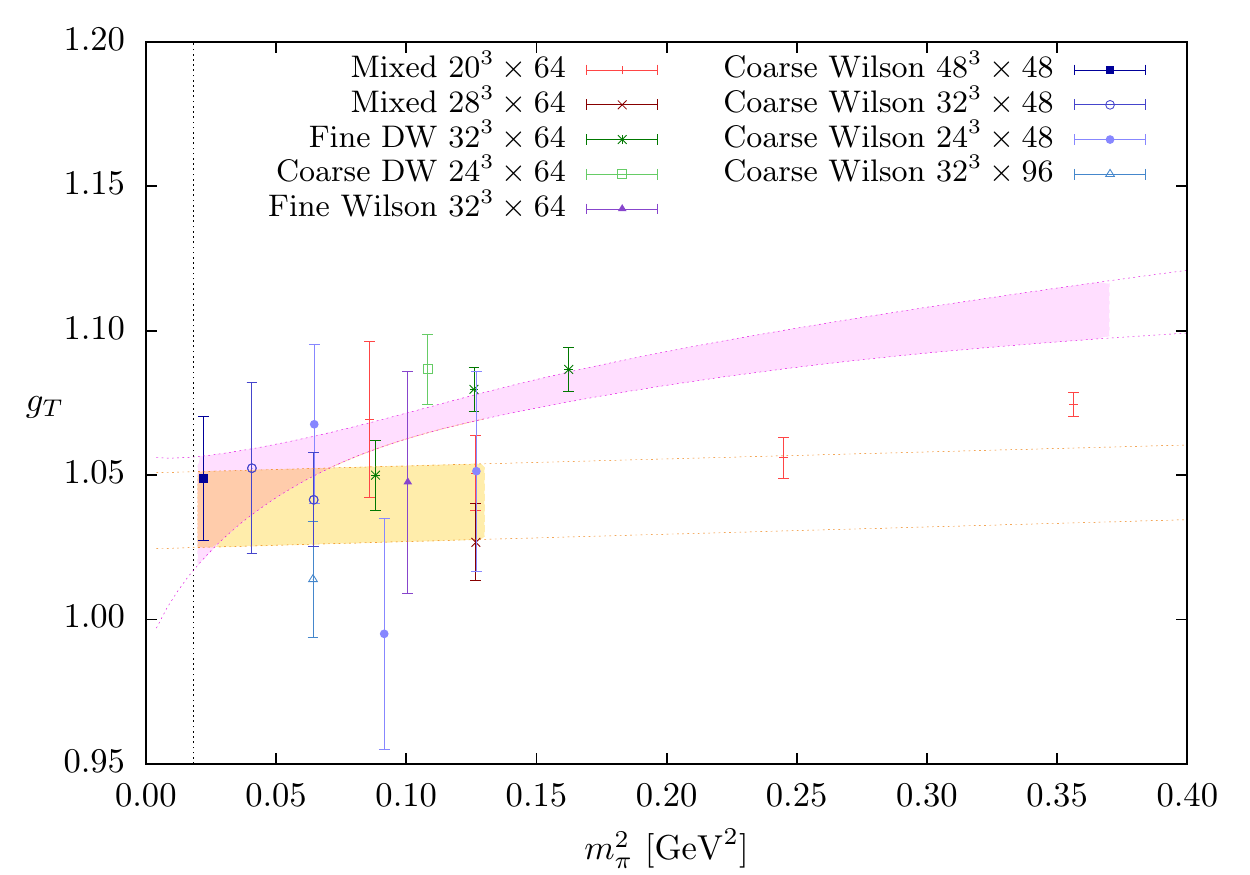}
  \caption{$g_T$ versus $m_\pi^2$. Error bars are purely statistical
    and do not include errors in renormalization factors, which are
    correlated across ensembles with the same action. Two chiral fits
    are shown: a two-parameter fit to the coarse Wilson-clover
    ensembles (orange) and a three-parameter fit to all shown ensembles
    (magenta).}
\label{fig:gTmpi}
\end{figure}

Our tensor charge results are shown in Fig.~\ref{fig:gTmpi}. Note that
the vertical scale is expanded and the error bars are much smaller. Again there is
general agreement between different lattice actions. The figure shows
some inconsistency between the domain wall and mixed-action ensembles
at larger pion masses. However, this is close to the 2\% correlated error
that we assign to the mixed-action results to account for perturbative
renormalization. As with the scalar charge, we estimate the physical
value of $g_T$ in three ways:
\begin{enumerate}
\item Using the value from the $m_\pi=149$~MeV ensemble yields $g_T=1.049(23)$.
\item Performing a 2-parameter chiral fit to the coarse Wilson-clover
  ensembles. As with the scalar charge, the fit works well with
  $\chi^2/\text{dof}=4.49/5$. However, the fit chooses $\mu^2=0$, which
  removes the chiral log. At the physical pion mass, this gives
  $g_T=1.038(11)$.
\item Performing a 3-parameter fit to the full set of
  ensembles. Because of the correlated error for the mixed-action
  ensembles, this fit also works well, giving
  $\chi^2/\text{dof}=19.69/14$. Extrapolating to the physical pion
  mass yields $g_T=1.037(20)$.
\end{enumerate}

%% file: conclusions.tex
Motivated by the facts that contributions of non-standard scalar and
tensor couplings to neutron decay are proportional to isovector $g_S$ and
$g_T$, and that $\bar{g}_{\pi N N}$, which produces parity-violating
components of nucleon-nucleon interactions, is proportional to isovector $g_S$,
we have undertaken a comprehensive set of calculations of $g_S$ and
$g_T$ using Lattice QCD.

Our calculation includes a number of significant advances.  We have
utilized three sets of ensembles that cover different pion mass regions for
which we use the same calculational methodology. The primary data for
this analysis were obtained using the smeared Wilson-clover action
developed by the BMW collaboration and include pion masses in the vicinity
of 150, 200, 250, 300, and 350~MeV. The 149~MeV pion mass plays a crucial
role in our analysis. It is so close to the physical pion mass that it
virtually eliminates uncertainty from extrapolation to the physical
mass and is well below the lowest mass used in other calculations.
Furthermore, given the known strong source-sink separation dependence in
some observables such as the momentum fraction, our careful study of
3 separations in Figs.~\ref{fig:gSsep} and Fig.~\ref{fig:gTsep} from 0.93~fm to
1.39~fm is important in ruling out
contributions of excited states near the physical pion mass.

To control systematic effects from finite lattice volume, we have
satisfied the standard criterion $m_\pi L_x a > 4$ for nearly all of
our ensembles, including the crucial $m_\pi = 149\, \mbox{MeV}$
ensemble where $m_\pi L_x a = 4.2$, cf.~Table~\ref{tab:ensembles}.  In
addition, we carried out an explicit test for finite volume effects
using the two ensembles W5 and W6 at $m_{\pi } =254\, \mbox{MeV} $,
which are identical except for their lattice volume. Within the
statistical accuracy of our calculation, no significant deviations
between the results obtained in these two ensembles were found.

Comparison of values of $g_S$ and $g_T$ for the ensembles W1 with
lattice spacing $a=0.09\, \mbox{fm}$ and W7 with $a=0.116\, \mbox{fm}$
at similar values of $m_\pi$ and $L_x a$, cf.~Table~\ref{tab:ensembles},
indicates the absence of $a$-dependence within our statistical
errors.

Supplementing the results obtained at the pion mass
$m_{\pi } =149\, \mbox{MeV} $, we
performed 3 and 2-parameter chiral fits to $g_S$ and $g_T$ for
the low mass Wilson ensembles, where we expect chiral perturbation
theory to be valid. These fits serve as a check on the consistency of
the 149~MeV data point and have the potential to reduce the
statistical error. The central values of the 149 MeV data points and
fits were essentially identical on the scale of the errors, and it
turned out that the error in the scalar charge was essentially
unchanged while the error on the tensor charge was cut in half. We
regard the chiral fits to the Wilson-clover data as our definitive
results for the central values and statistical error.
We also estimate the systematic error from excited-state contamination
as the average of two absolute differences; those between the central value
from the middle source-sink separation on the 149~MeV ensemble and the
central values from the other two source-sink separations on that
ensemble, i.e.,
$\tfrac{1}{2}(|g_X^{T=10a}-g_X^{T=8a}|+|g_X^{T=10a}-g_X^{T=12a}|)$.
This yields $g_S = 1.08 \pm 0.28 \text{ (stat)} \pm 0.16 \text{ (syst)}$
and $g_T = 1.038 \pm 0.011 \text{ (stat)} \pm 0.012 \text{ (syst)}$.
We emphasize that our results do not hinge decisively on chiral perturbation
theory. The chiral fits to the data at higher pion masses rather serve to
buttress the calculations at $m_{\pi } =149\, \mbox{MeV} $, which by
themselves represent determinations of $g_S $ and $g_T $ with essentially
no residual uncertainty due to chiral extrapolation.

The range of pion masses for the Wilson action overlaps the range 297
to 403 MeV for the domain wall action and the range 293 to 597 MeV for
the mixed action, so simultaneous analysis provides a valuable
consistency check of the actions and normalization. In the region of
overlap between the Wilson-clover and domain wall calculations, the
results agreed well, confirming the underlying consistency of these
two calculations including normalization. There is a certain tension
between the mixed action and domain wall results at higher pion
masses, the origin of which we have not been able to pinpoint; an
underestimate of the renormalization uncertainties in the mixed action
case is one possible source. When the error in the normalization is
taken into account, the 4 and 3-parameter chiral fits to $g_S$ and
$g_T$ for all three sets of ensembles are consistent with the chiral fits to
the low mass Wilson ensembles, as reflected in the close agreement of
the error envelopes at the physical pion mass.

Concerning other systematic uncertainties, we have yet to perform extensive
low mass Wilson calculations at smaller lattice spacings to
extrapolate to the continuum limit  and at larger volumes to
extrapolate to the infinite volume limit.  At the present level of
statistics, our comparison of two different lattice volumes at the
pion mass $m_\pi= 254\,\mbox{MeV}$ and two different lattice spacings at $m_\pi \approx
300\,\mbox{MeV}$ yielded no evidence of significant volume or lattice spacing
effects. However, in view of past experience with the well-studied
case of the nucleon axial charge $g_A$, future more statistically
accurate calculations may yet necessitate more detailed investigations
of a range of volumes and lattice spacings.  Comparable calculations
of $g_A$ by several groups including our own lie below the experimental
value by more than the statistical errors, and it is likewise possible
that the aforementioned or other systematic errors in $g_S$ and $g_T$ are
still appreciable.

%% file: app-npr.tex
\begin{figure}[ht!]
  \centering
  \includegraphics[width=.49\textwidth]{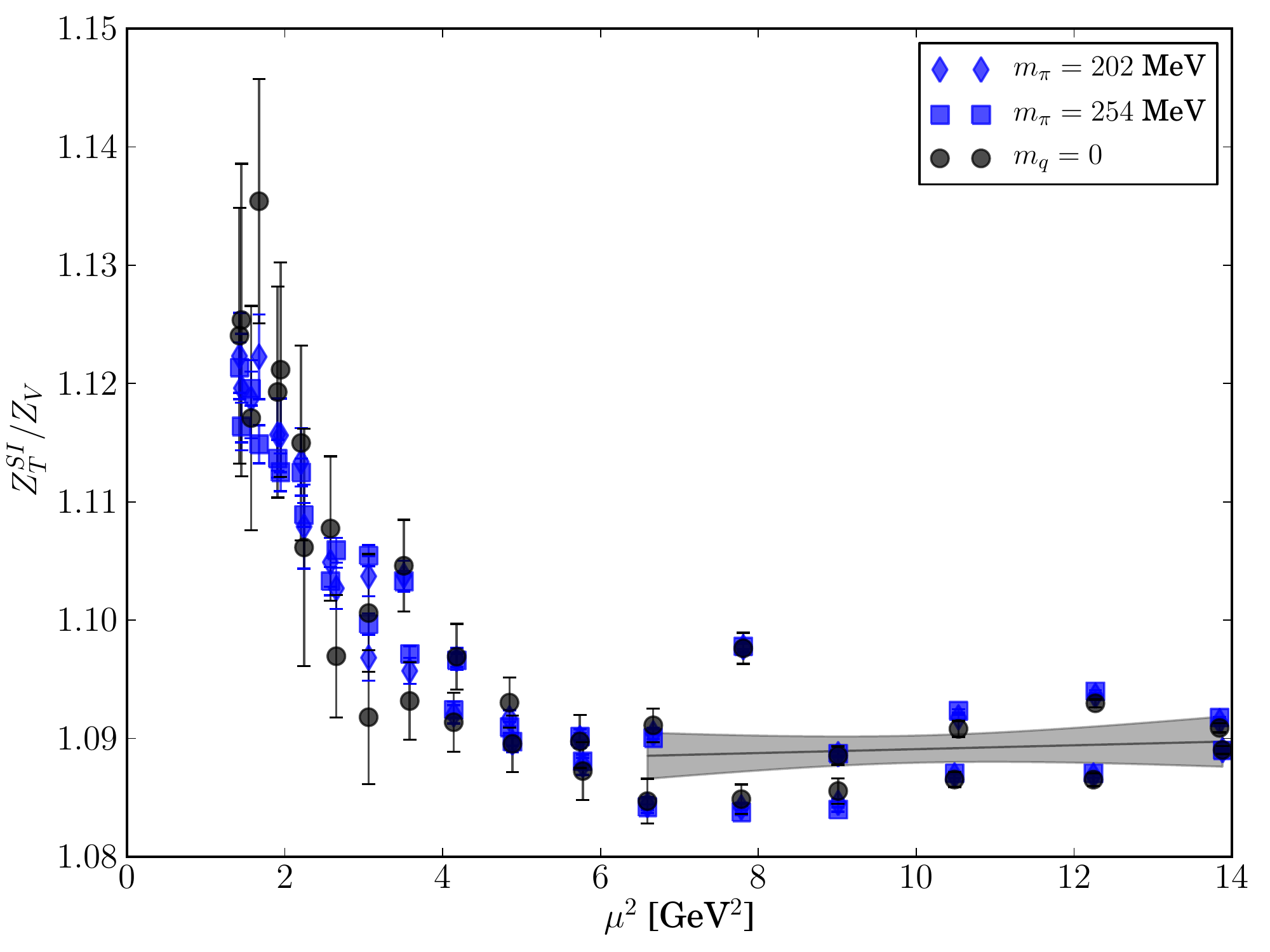}~
  \includegraphics[width=.49\textwidth]{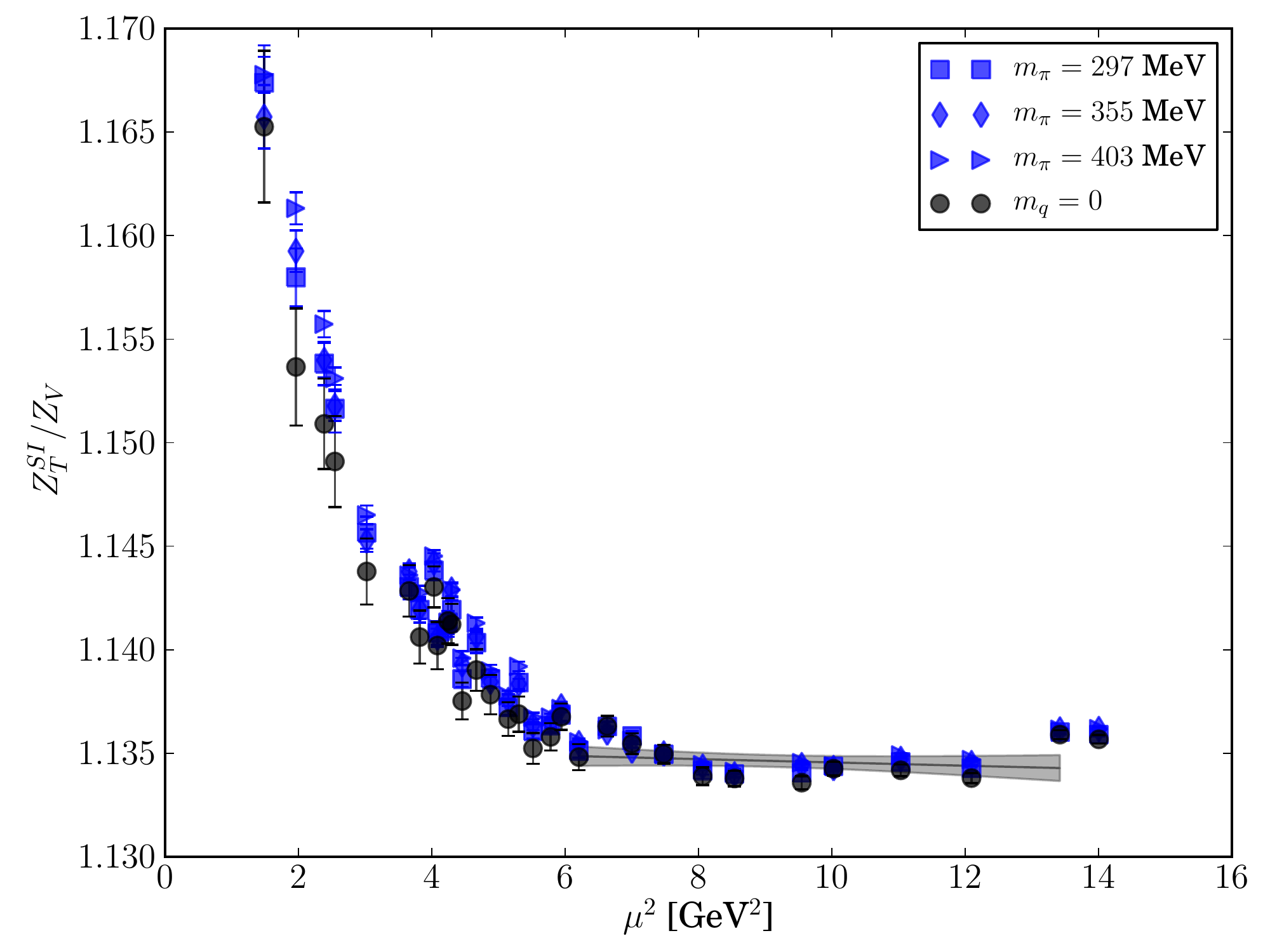}
  \caption{\label{fig:npr-zT}
    Extrapolations of the scale-independent (SI) ratio $\big(Z_T^\text{SI}/Z_V^\text{lat}\big)$ 
    for coarse clover-improved Wilson fermion lattice ensembles (W3, W5) (left) and
    fine domain wall fermion lattices (D1, D2, D3) (right).}
\end{figure}

In this section, we briefly describe the method and present the results of computing the
nonperturbative renormalization factors $Z_T^{\overline{MS}(2\text{ GeV})}$
for the tensor charge $g_T$, 
$Z_T^{\overline{MS}(2\text{ GeV})} g_T^\text{lat} = g_T^{\overline{MS}(2\text{ GeV})}$,
for clover-improved Wilson and domain wall fermion ensembles.

Following Ref.~\cite{Martinelli:1994ty}, we compute amputated lattice Green functions 
$\Pi_T^\text{lat}$ of tensor charge operator and plane wave quarks 
$q(x)\sim e^{ipx}$, $p^2=\mu^2$ in the Landau gauge.
We then extract the lattice renormalization factor $Z_T^\text{lat}$ defined as 
$(Z / Z_q)^\text{lat} \Pi^\text{lat} = \Pi^\text{tree} $, 
where $\Pi^\text{tree}$ is the corresponding tree-level amputated Green function
and $Z_q$ is the quark field renormalization.
In order to match lattice-regulated and perturbatively renormalized operators, 
we compute scale-independent (SI) renormalization factors $Z_T^\text{SI}$
\begin{equation}
\frac{Z_T^\text{SI}}{Z_V} 
  = \Big[\Big(\frac{Z_T}{Z_V}\Big)^\text{lat}_{p^2=\mu^2}
         \cdot \Big(\frac{Z_T^{RI'(2\text{ GeV})}}{Z_T^{RI'(\mu)}}\Big)^\text{pert}
    \Big]_{m_q\to0,\,\mu\to0}\,,
\end{equation}
where the perturbative operator renormalization is calculated to order 
$O(\alpha_S^3)$~\cite{Gracey:2003yr}
and $Z_V^\text{lat}$ is the renormalization of the quark charge operator on the lattice,
\begin{equation}
(Z_V / Z_q)^\text{lat} \cdot \langle N(\vec P)| u^\dag u - d^\dag d | N(\vec P) \rangle_\text{lat}
  \doteq 1\,.
\end{equation}
We extrapolate linearly first in $m_q\to0$, and then in $\mu^2\to0$ 
in the range $\mu^2\approx6\ldots15\text{ GeV}^2$, see Fig.~\ref{fig:npr-zT}.
Finally, we obtain the renormalization factors for converting lattice results 
into the $\overline{MS}(2\text{ GeV})$ scheme:
\begin{equation}
Z^{\overline{MS}(2\text{ GeV})}_T 
  = \Big(\frac{Z_T^\text{SI}}{Z_V}\Big)\cdot Z_V^\text{lat}\cdot 
    \Big(\frac{Z_T^{\overline{MS}(2\text{ GeV})}}{Z_T^{RI'(2\text{ GeV})}}\Big)^\text{pert}\,.
\end{equation}

%% file: chipt_scalar.tex
For chiral extrapolation of the scalar charge, we make use of the
Feynman-Hellmann theorem to obtain
\begin{equation}
  g_S \equiv \langle p | \bar u u - \bar d d | p \rangle =
  \left( \frac{\partial}{\partial m_u} - \frac{\partial}{\partial m_d}
  \right) M_p.
\end{equation}
Strong isospin splitting of nucleon masses has been
computed in $SU(2)$ heavy baryon chiral perturbation theory, with
explicit $\Delta$ degrees of freedom, to
next-to-next-to-leading-order, i.e.\ to $\mathcal{O}(m_q^2)$
\cite{Tiburzi:2005na}. This leads to the expression
\begin{equation}
  \begin{split}
    g_S(m_\pi^2) &= 2\alpha_M - \frac{1}{2(4\pi F_\pi)^2}\Biggl\{ m_\pi^2\left[8(g_A^2\alpha_M + g_{\Delta N}^2(\alpha_M + \tfrac{5}{9}\gamma_M)) - (b_1^M+b_6^M)\frac{\pi (\sqrt{2} F_\pi)^3}{\lambda}\right] \\
    &\qquad +m_\pi^2\log\left(\frac{m_\pi^2}{\mu^2}\right)2\alpha_M(6 g_A^2 + 1) 
    + \mathcal{J}(m_\pi,\Delta,\mu) 8g_{\Delta N}^2(\alpha_M + \tfrac{5}{9}\gamma_M) \Biggr\},
  \end{split}
\end{equation}
where
\begin{equation}
\mathcal{J}(m,\Delta,\mu) = (m^2-2\Delta^2)\log\frac{m^2}{\mu^2} + 2\Delta\sqrt{\Delta^2-m^2}\log\left(\frac{\Delta - \sqrt{\Delta^2-m^2+i\epsilon}}{\Delta + \sqrt{\Delta^2-m^2+i\epsilon}}\right) + 2\Delta^2\log\frac{4\Delta^2}{\mu^2}.
\end{equation}
Here $F_\pi$, $g_A$, and $\Delta$ are the pion decay constant, the
nucleon axial charge, and the $\Delta(1232)$-nucleon mass splitting,
respectively, in the $SU(2)$ chiral limit. The convention used here is
that the physical value of $F_\pi$ is approximately 92~MeV. We fix
these three parameters to values used for chiral extrapolation in a
previous nucleon structure calculation \cite{Bratt:2010jn},
\begin{equation}
  F_\pi=86.2\text{ MeV}, \qquad g_A=1.2, \qquad \Delta=0.2711\text{ GeV},
\end{equation}
leaving 3 independent parameters to which we fit our
$g_S(m_\pi^2)$ data. If we fix only $\Delta$, then the resulting
fit has 4 independent parameters.

%% file: chipt_tensor.tex
Chiral extrapolation formulas for the tensor charge are given in
\cite{Detmold:2002nf}. Full formulas including $\Delta$ loops are
given via a number of integrals, and then it is shown that these are
approximated well by an expression that includes only the leading
non-analytic term. Also including terms that connect to the heavy
quark limit:
\begin{equation}
  g_T(m_\pi^2) = \delta a\left(1 
     + \delta c_\text{LNA} m_\pi^2 \log\frac{m_\pi^2}{m_\pi^2 + \mu^2}\right) 
   + \delta b \frac{m_\pi^2}{m_\pi^2+m_b^2},
\end{equation}
where from the heavy quark limit
\begin{equation}
 \delta b = \tfrac{5}{3} - \delta a (1 - \mu^2\delta c_\text{LNA}),
\end{equation}
and the coefficient of the $\log$ term:
\begin{equation}
 \delta c_\text{LNA} = \frac{-1}{2(4\pi F_\pi)^2}\left[\left(2-\frac{4}{9}\frac{g^2_{\pi N \Delta}}{g^2_{\pi NN}}\right)g_A^2 + \frac{1}{2}\right].
\end{equation}
Fixing $m_b=5$~GeV as in \cite{Detmold:2002nf} yields a fit with 3
independent parameters. In addition, fixing $g_A$ and $F_\pi$ as for
the scalar charge, and $\frac{g_{\pi N\Delta}}{g_{\pi NN}} = 1.85$,
reduces the number of independent parameters to 2.

%% file: table.tex
Table~\ref{tab:results} contains the scalar and tensor data from our
full set of ensembles.

\begin{table}
  \centering
  \begin{tabular}{c|D{?}{}{5.5}|D{.}{.}{3.8}D{.}{.}{4.9}|D{.}{.}{3.7}D{.}{.}{3.7}}
    && \multicolumn{2}{c|}{Scalar} & \multicolumn{2}{c}{Tensor} \\
    Ensemble & \multicolumn{1}{c|}{$m_\pi$ [MeV]} &\multicolumn{1}{c}{$u-d$} & \multicolumn{1}{c|}{$u+d$} & \multicolumn{1}{c}{$u-d$} & \multicolumn{1}{c}{$u+d$} \\\hline
    W1 & 317?(2) & 1.017(248) & 5.618(388) & 1.047(38) & 0.700(46) \\
    W2 & 149?(1) & 1.006(271) & 10.127(1.501) & 1.049(22) & 0.652(33) \\
    W3 & 202?(1) & 1.009(235) & 6.632(620) & 1.052(30) & 0.622(46) \\
    W4 & 253?(1) & 1.335(262) & 7.808(906) & 1.014(20) & 0.598(55) \\
    W5 & 254?(1) & 0.901(110) & 6.027(251) & 1.041(16) & 0.676(22) \\
    W6 & 254?(1) & 0.757(195) & 5.241(935) & 1.068(28) & 0.668(33) \\
    W7 & 303?(2) & 0.622(260) & 4.923(395) & 0.995(40) & 0.656(63) \\
    W8 & 356?(2) & 0.817(168) & 5.128(268) & 1.051(35) & 0.691(46) \\
    D1 & 297?(5) & 0.964(85) & 8.933(195) & 1.050(12) & 0.779(19) \\
    D2 & 355?(6) & 0.854(43) & 8.114(102) & 1.080(8) & 0.807(11) \\
    D3 & 403?(7) & 0.941(35) & 7.903(81) & 1.087(8) & 0.802(11) \\
    D4 & 329?(5) & 0.839(65) & 8.193(153) & 1.087(12) & 0.827(18) \\
    M1 & 293?(6) & 1.012(208) & 6.624(278) & 1.069(27) & 0.543(34) \\
    M2 & 356?(7) & 0.624(73) & 4.912(89) & 1.027(13) & 0.546(14) \\
    M3 & 356?(7) & 0.648(67) & 4.944(97) & 1.051(13) & 0.557(14) \\
    M4 & 495?(10)& 0.787(25) & 3.908(35) & 1.056(7) & 0.566(7) \\
    M5 & 597?(12)&  0.814(13) & 3.495(17) & 1.074(4) & 0.574(4)
  \end{tabular}
  \caption{Full set of results for the scalar and tensor charges. The isovector data are renormalized in the $\overline{MS}$ scheme at $\mu=2$~GeV, whereas the $u+d$ data contain only the connected contribution and are not renormalized. Wilson data, W1--W8, are from the middle source-sink separation.}
  \label{tab:results}
\end{table}

%% file: paper.bbl
\begin{thebibliography}{33}%
\makeatletter
\providecommand \@ifxundefined [1]{%
 \@ifx{#1\undefined}
}%
\providecommand \@ifnum [1]{%
 \ifnum #1\expandafter \@firstoftwo
 \else \expandafter \@secondoftwo
 \fi
}%
\providecommand \@ifx [1]{%
 \ifx #1\expandafter \@firstoftwo
 \else \expandafter \@secondoftwo
 \fi
}%
\providecommand \natexlab [1]{#1}%
\providecommand \enquote  [1]{``#1''}%
\providecommand \bibnamefont  [1]{#1}%
\providecommand \bibfnamefont [1]{#1}%
\providecommand \citenamefont [1]{#1}%
\providecommand \href@noop [0]{\@secondoftwo}%
\providecommand \href [0]{\begingroup \@sanitize@url \@href}%
\providecommand \@href[1]{\@@startlink{#1}\@@href}%
\providecommand \@@href[1]{\endgroup#1\@@endlink}%
\providecommand \@sanitize@url [0]{\catcode `\\12\catcode `\$12\catcode
  `\&12\catcode `\#12\catcode `\^12\catcode `\_12\catcode `\%12\relax}%
\providecommand \@@startlink[1]{}%
\providecommand \@@endlink[0]{}%
\providecommand \url  [0]{\begingroup\@sanitize@url \@url }%
\providecommand \@url [1]{\endgroup\@href {#1}{\urlprefix }}%
\providecommand \urlprefix  [0]{URL }%
\providecommand \Eprint [0]{\href }%
\providecommand \doibase [0]{http://dx.doi.org/}%
\providecommand \selectlanguage [0]{\@gobble}%
\providecommand \bibinfo  [0]{\@secondoftwo}%
\providecommand \bibfield  [0]{\@secondoftwo}%
\providecommand \translation [1]{[#1]}%
\providecommand \BibitemOpen [0]{}%
\providecommand \bibitemStop [0]{}%
\providecommand \bibitemNoStop [0]{.\EOS\space}%
\providecommand \EOS [0]{\spacefactor3000\relax}%
\providecommand \BibitemShut  [1]{\csname bibitem#1\endcsname}%
\let\auto@bib@innerbib\@empty
\bibitem [{\citenamefont {Bhattacharya}\ \emph {et~al.}(2012)\citenamefont
  {Bhattacharya}, \citenamefont {Cirigliano}, \citenamefont {Cohen},
  \citenamefont {Filipuzzi}, \citenamefont {Gonzalez-Alonso} \emph
  {et~al.}}]{Bhattacharya:2011qm}%
  \BibitemOpen
  \bibfield  {author} {\bibinfo {author} {\bibfnamefont {T.}~\bibnamefont
  {Bhattacharya}}, \bibinfo {author} {\bibfnamefont {V.}~\bibnamefont
  {Cirigliano}}, \bibinfo {author} {\bibfnamefont {S.~D.}\ \bibnamefont
  {Cohen}}, \bibinfo {author} {\bibfnamefont {A.}~\bibnamefont {Filipuzzi}},
  \bibinfo {author} {\bibfnamefont {M.}~\bibnamefont {Gonzalez-Alonso}},  \emph
  {et~al.},\ }\href {\doibase 10.1103/PhysRevD.85.054512} {\bibfield  {journal}
  {\bibinfo  {journal} {Phys.Rev.}\ }\textbf {\bibinfo {volume} {D85}},\
  \bibinfo {pages} {054512} (\bibinfo {year} {2012})},\ \Eprint
  {http://arxiv.org/abs/1110.6448} {arXiv:1110.6448 [hep-ph]} \BibitemShut
  {NoStop}%
\bibitem [{\citenamefont {Shintani}\ \emph {et~al.}(2007)\citenamefont
  {Shintani}, \citenamefont {Aoki}, \citenamefont {Ishizuka}, \citenamefont
  {Kanaya}, \citenamefont {Kikukawa} \emph {et~al.}}]{Shintani:2006xr}%
  \BibitemOpen
  \bibfield  {author} {\bibinfo {author} {\bibfnamefont {E.}~\bibnamefont
  {Shintani}}, \bibinfo {author} {\bibfnamefont {S.}~\bibnamefont {Aoki}},
  \bibinfo {author} {\bibfnamefont {N.}~\bibnamefont {Ishizuka}}, \bibinfo
  {author} {\bibfnamefont {K.}~\bibnamefont {Kanaya}}, \bibinfo {author}
  {\bibfnamefont {Y.}~\bibnamefont {Kikukawa}},  \emph {et~al.},\ }\href
  {\doibase 10.1103/PhysRevD.75.034507} {\bibfield  {journal} {\bibinfo
  {journal} {Phys.Rev.}\ }\textbf {\bibinfo {volume} {D75}},\ \bibinfo {pages}
  {034507} (\bibinfo {year} {2007})},\ \Eprint
  {http://arxiv.org/abs/hep-lat/0611032} {arXiv:hep-lat/0611032 [hep-lat]}
  \BibitemShut {NoStop}%
\bibitem [{\citenamefont {Shintani}\ \emph {et~al.}(2008)\citenamefont
  {Shintani}, \citenamefont {Aoki},\ and\ \citenamefont
  {Kuramashi}}]{Shintani:2008nt}%
  \BibitemOpen
  \bibfield  {author} {\bibinfo {author} {\bibfnamefont {E.}~\bibnamefont
  {Shintani}}, \bibinfo {author} {\bibfnamefont {S.}~\bibnamefont {Aoki}}, \
  and\ \bibinfo {author} {\bibfnamefont {Y.}~\bibnamefont {Kuramashi}},\ }\href
  {\doibase 10.1103/PhysRevD.78.014503} {\bibfield  {journal} {\bibinfo
  {journal} {Phys.Rev.}\ }\textbf {\bibinfo {volume} {D78}},\ \bibinfo {pages}
  {014503} (\bibinfo {year} {2008})},\ \Eprint {http://arxiv.org/abs/0803.0797}
  {arXiv:0803.0797 [hep-lat]} \BibitemShut {NoStop}%
\bibitem [{\citenamefont {Aoki}\ \emph {et~al.}(2008)\citenamefont {Aoki},
  \citenamefont {Horsley}, \citenamefont {Izubuchi}, \citenamefont {Nakamura},
  \citenamefont {Pleiter} \emph {et~al.}}]{Aoki:2008gv}%
  \BibitemOpen
  \bibfield  {author} {\bibinfo {author} {\bibfnamefont {S.}~\bibnamefont
  {Aoki}}, \bibinfo {author} {\bibfnamefont {R.}~\bibnamefont {Horsley}},
  \bibinfo {author} {\bibfnamefont {T.}~\bibnamefont {Izubuchi}}, \bibinfo
  {author} {\bibfnamefont {Y.}~\bibnamefont {Nakamura}}, \bibinfo {author}
  {\bibfnamefont {D.}~\bibnamefont {Pleiter}},  \emph {et~al.},\ }\href@noop {}
  {\  (\bibinfo {year} {2008})},\ \Eprint {http://arxiv.org/abs/0808.1428}
  {arXiv:0808.1428 [hep-lat]} \BibitemShut {NoStop}%
\bibitem [{\citenamefont {Liu}(2009)}]{Liu:2008gr}%
  \BibitemOpen
  \bibfield  {author} {\bibinfo {author} {\bibfnamefont {K.-F.}\ \bibnamefont
  {Liu}},\ }\href {\doibase 10.1142/S0217732309031375} {\bibfield  {journal}
  {\bibinfo  {journal} {Mod.Phys.Lett.}\ }\textbf {\bibinfo {volume} {A24}},\
  \bibinfo {pages} {1971} (\bibinfo {year} {2009})},\ \Eprint
  {http://arxiv.org/abs/0807.1365} {arXiv:0807.1365 [hep-ph]} \BibitemShut
  {NoStop}%
\bibitem [{\citenamefont {Crewther}\ \emph {et~al.}(1979)\citenamefont
  {Crewther}, \citenamefont {Di~Vecchia}, \citenamefont {Veneziano},\ and\
  \citenamefont {Witten}}]{Crewther:1979pi}%
  \BibitemOpen
  \bibfield  {author} {\bibinfo {author} {\bibfnamefont {R.}~\bibnamefont
  {Crewther}}, \bibinfo {author} {\bibfnamefont {P.}~\bibnamefont
  {Di~Vecchia}}, \bibinfo {author} {\bibfnamefont {G.}~\bibnamefont
  {Veneziano}}, \ and\ \bibinfo {author} {\bibfnamefont {E.}~\bibnamefont
  {Witten}},\ }\href {\doibase 10.1016/0370-2693(79)90128-X,
  10.1016/0370-2693(79)90128-X} {\bibfield  {journal} {\bibinfo  {journal}
  {Phys.Lett.}\ }\textbf {\bibinfo {volume} {B88}},\ \bibinfo {pages} {123}
  (\bibinfo {year} {1979})}\BibitemShut {NoStop}%
\bibitem [{\citenamefont {Haxton}\ and\ \citenamefont
  {Henley}(1983)}]{Haxton:1983dq}%
  \BibitemOpen
  \bibfield  {author} {\bibinfo {author} {\bibfnamefont {W.}~\bibnamefont
  {Haxton}}\ and\ \bibinfo {author} {\bibfnamefont {E.}~\bibnamefont
  {Henley}},\ }\href {\doibase 10.1103/PhysRevLett.51.1937} {\bibfield
  {journal} {\bibinfo  {journal} {Phys.Rev.Lett.}\ }\textbf {\bibinfo {volume}
  {51}},\ \bibinfo {pages} {1937} (\bibinfo {year} {1983})}\BibitemShut
  {NoStop}%
\bibitem [{\citenamefont {Ohki}\ \emph {et~al.}(2008)\citenamefont {Ohki},
  \citenamefont {Fukaya}, \citenamefont {Hashimoto}, \citenamefont {Kaneko},
  \citenamefont {Matsufuru} \emph {et~al.}}]{Ohki:2008ff}%
  \BibitemOpen
  \bibfield  {author} {\bibinfo {author} {\bibfnamefont {H.}~\bibnamefont
  {Ohki}}, \bibinfo {author} {\bibfnamefont {H.}~\bibnamefont {Fukaya}},
  \bibinfo {author} {\bibfnamefont {S.}~\bibnamefont {Hashimoto}}, \bibinfo
  {author} {\bibfnamefont {T.}~\bibnamefont {Kaneko}}, \bibinfo {author}
  {\bibfnamefont {H.}~\bibnamefont {Matsufuru}},  \emph {et~al.},\ }\href
  {\doibase 10.1103/PhysRevD.78.054502} {\bibfield  {journal} {\bibinfo
  {journal} {Phys.Rev.}\ }\textbf {\bibinfo {volume} {D78}},\ \bibinfo {pages}
  {054502} (\bibinfo {year} {2008})},\ \Eprint {http://arxiv.org/abs/0806.4744}
  {arXiv:0806.4744 [hep-lat]} \BibitemShut {NoStop}%
\bibitem [{\citenamefont {D\"urr}\ \emph {et~al.}(2012)\citenamefont {D\"urr},
  \citenamefont {Fodor}, \citenamefont {Hemmert}, \citenamefont {Hoelbling},
  \citenamefont {Frison} \emph {et~al.}}]{Durr:2011mp}%
  \BibitemOpen
  \bibfield  {author} {\bibinfo {author} {\bibfnamefont {S.}~\bibnamefont
  {D\"urr}}, \bibinfo {author} {\bibfnamefont {Z.}~\bibnamefont {Fodor}},
  \bibinfo {author} {\bibfnamefont {T.}~\bibnamefont {Hemmert}}, \bibinfo
  {author} {\bibfnamefont {C.}~\bibnamefont {Hoelbling}}, \bibinfo {author}
  {\bibfnamefont {J.}~\bibnamefont {Frison}},  \emph {et~al.},\ }\href
  {\doibase 10.1103/PhysRevD.85.014509} {\bibfield  {journal} {\bibinfo
  {journal} {Phys.Rev.}\ }\textbf {\bibinfo {volume} {D85}},\ \bibinfo {pages}
  {014509} (\bibinfo {year} {2012})},\ \Eprint {http://arxiv.org/abs/1109.4265}
  {arXiv:1109.4265 [hep-lat]} \BibitemShut {NoStop}%
\bibitem [{\citenamefont {Horsley}\ \emph {et~al.}(2012)\citenamefont
  {Horsley}, \citenamefont {Nakamura}, \citenamefont {Perlt}, \citenamefont
  {Pleiter}, \citenamefont {Rakow} \emph {et~al.}}]{Horsley:2011wr}%
  \BibitemOpen
  \bibfield  {author} {\bibinfo {author} {\bibfnamefont {R.}~\bibnamefont
  {Horsley}}, \bibinfo {author} {\bibfnamefont {Y.}~\bibnamefont {Nakamura}},
  \bibinfo {author} {\bibfnamefont {H.}~\bibnamefont {Perlt}}, \bibinfo
  {author} {\bibfnamefont {D.}~\bibnamefont {Pleiter}}, \bibinfo {author}
  {\bibfnamefont {P.}~\bibnamefont {Rakow}},  \emph {et~al.},\ }\href {\doibase
  10.1103/PhysRevD.85.034506} {\bibfield  {journal} {\bibinfo  {journal}
  {Phys.Rev.}\ }\textbf {\bibinfo {volume} {D85}},\ \bibinfo {pages} {034506}
  (\bibinfo {year} {2012})},\ \Eprint {http://arxiv.org/abs/1110.4971}
  {arXiv:1110.4971 [hep-lat]} \BibitemShut {NoStop}%
\bibitem [{\citenamefont {Bottino}\ \emph {et~al.}(2002)\citenamefont
  {Bottino}, \citenamefont {Donato}, \citenamefont {Fornengo},\ and\
  \citenamefont {Scopel}}]{Bottino:2001dj}%
  \BibitemOpen
  \bibfield  {author} {\bibinfo {author} {\bibfnamefont {A.}~\bibnamefont
  {Bottino}}, \bibinfo {author} {\bibfnamefont {F.}~\bibnamefont {Donato}},
  \bibinfo {author} {\bibfnamefont {N.}~\bibnamefont {Fornengo}}, \ and\
  \bibinfo {author} {\bibfnamefont {S.}~\bibnamefont {Scopel}},\ }\href
  {\doibase 10.1016/S0927-6505(02)00107-X} {\bibfield  {journal} {\bibinfo
  {journal} {Astropart.Phys.}\ }\textbf {\bibinfo {volume} {18}},\ \bibinfo
  {pages} {205} (\bibinfo {year} {2002})},\ \Eprint
  {http://arxiv.org/abs/hep-ph/0111229} {arXiv:hep-ph/0111229 [hep-ph]}
  \BibitemShut {NoStop}%
\bibitem [{\citenamefont {Ellis}\ \emph {et~al.}(2005)\citenamefont {Ellis},
  \citenamefont {Olive}, \citenamefont {Santoso},\ and\ \citenamefont
  {Spanos}}]{Ellis:2005mb}%
  \BibitemOpen
  \bibfield  {author} {\bibinfo {author} {\bibfnamefont {J.~R.}\ \bibnamefont
  {Ellis}}, \bibinfo {author} {\bibfnamefont {K.~A.}\ \bibnamefont {Olive}},
  \bibinfo {author} {\bibfnamefont {Y.}~\bibnamefont {Santoso}}, \ and\
  \bibinfo {author} {\bibfnamefont {V.~C.}\ \bibnamefont {Spanos}},\ }\href
  {\doibase 10.1103/PhysRevD.71.095007} {\bibfield  {journal} {\bibinfo
  {journal} {Phys.Rev.}\ }\textbf {\bibinfo {volume} {D71}},\ \bibinfo {pages}
  {095007} (\bibinfo {year} {2005})},\ \Eprint
  {http://arxiv.org/abs/hep-ph/0502001} {arXiv:hep-ph/0502001 [hep-ph]}
  \BibitemShut {NoStop}%
\bibitem [{\citenamefont {Bertone}(2010)}]{Bertone:2010zz}%
  \BibitemOpen
  \bibinfo {editor} {\bibfnamefont {G.}~\bibnamefont {Bertone}},\ ed.,\
  \href@noop {} {\emph {\bibinfo {title} {{Particle dark matter: Observations,
  models and searches}}}}\ (\bibinfo  {publisher} {Cambridge University
  Press},\ \bibinfo {year} {2010})\BibitemShut {NoStop}%
\bibitem [{\citenamefont {D\"urr}\ \emph
  {et~al.}(2011{\natexlab{a}})\citenamefont {D\"urr} \emph
  {et~al.}}]{Durr:2010aw}%
  \BibitemOpen
  \bibfield  {author} {\bibinfo {author} {\bibfnamefont {S.}~\bibnamefont
  {D\"urr}} \emph {et~al.},\ }\href {\doibase 10.1007/JHEP08(2011)148}
  {\bibfield  {journal} {\bibinfo  {journal} {JHEP}\ }\textbf {\bibinfo
  {volume} {08}},\ \bibinfo {pages} {148} (\bibinfo {year}
  {2011}{\natexlab{a}})},\ \Eprint {http://arxiv.org/abs/1011.2711}
  {arXiv:1011.2711 [hep-lat]} \BibitemShut {NoStop}%
\bibitem [{\citenamefont {Allton}\ \emph {et~al.}(2008)\citenamefont {Allton}
  \emph {et~al.}}]{Allton:2008pn}%
  \BibitemOpen
  \bibfield  {author} {\bibinfo {author} {\bibfnamefont {C.}~\bibnamefont
  {Allton}} \emph {et~al.} (\bibinfo {collaboration} {RBC-UKQCD
  Collaboration}),\ }\href {\doibase 10.1103/PhysRevD.78.114509} {\bibfield
  {journal} {\bibinfo  {journal} {Phys.Rev.}\ }\textbf {\bibinfo {volume}
  {D78}},\ \bibinfo {pages} {114509} (\bibinfo {year} {2008})},\ \Eprint
  {http://arxiv.org/abs/0804.0473} {arXiv:0804.0473 [hep-lat]} \BibitemShut
  {NoStop}%
\bibitem [{\citenamefont {Aoki}\ \emph {et~al.}(2011)\citenamefont {Aoki} \emph
  {et~al.}}]{Aoki:2010dy}%
  \BibitemOpen
  \bibfield  {author} {\bibinfo {author} {\bibfnamefont {Y.}~\bibnamefont
  {Aoki}} \emph {et~al.} (\bibinfo {collaboration} {RBC Collaboration, UKQCD
  Collaboration}),\ }\href {\doibase 10.1103/PhysRevD.83.074508} {\bibfield
  {journal} {\bibinfo  {journal} {Phys.Rev.}\ }\textbf {\bibinfo {volume}
  {D83}},\ \bibinfo {pages} {074508} (\bibinfo {year} {2011})},\ \Eprint
  {http://arxiv.org/abs/1011.0892} {arXiv:1011.0892 [hep-lat]} \BibitemShut
  {NoStop}%
\bibitem [{\citenamefont {Syritsyn}\ \emph {et~al.}(2010)\citenamefont
  {Syritsyn}, \citenamefont {Bratt}, \citenamefont {Lin}, \citenamefont
  {Meyer}, \citenamefont {Negele} \emph {et~al.}}]{Syritsyn:2009mx}%
  \BibitemOpen
  \bibfield  {author} {\bibinfo {author} {\bibfnamefont {S.~N.}\ \bibnamefont
  {Syritsyn}}, \bibinfo {author} {\bibfnamefont {J.~D.}\ \bibnamefont {Bratt}},
  \bibinfo {author} {\bibfnamefont {M.~F.}\ \bibnamefont {Lin}}, \bibinfo
  {author} {\bibfnamefont {H.~B.}\ \bibnamefont {Meyer}}, \bibinfo {author}
  {\bibfnamefont {J.~W.}\ \bibnamefont {Negele}},  \emph {et~al.},\ }\href
  {\doibase 10.1103/PhysRevD.81.034507} {\bibfield  {journal} {\bibinfo
  {journal} {Phys.Rev.}\ }\textbf {\bibinfo {volume} {D81}},\ \bibinfo {pages}
  {034507} (\bibinfo {year} {2010})},\ \Eprint {http://arxiv.org/abs/0907.4194}
  {arXiv:0907.4194 [hep-lat]} \BibitemShut {NoStop}%
\bibitem [{\citenamefont {Aoki}\ \emph {et~al.}(2010)\citenamefont {Aoki},
  \citenamefont {Blum}, \citenamefont {Lin}, \citenamefont {Ohta},
  \citenamefont {Sasaki} \emph {et~al.}}]{Aoki:2010xg}%
  \BibitemOpen
  \bibfield  {author} {\bibinfo {author} {\bibfnamefont {Y.}~\bibnamefont
  {Aoki}}, \bibinfo {author} {\bibfnamefont {T.}~\bibnamefont {Blum}}, \bibinfo
  {author} {\bibfnamefont {H.-W.}\ \bibnamefont {Lin}}, \bibinfo {author}
  {\bibfnamefont {S.}~\bibnamefont {Ohta}}, \bibinfo {author} {\bibfnamefont
  {S.}~\bibnamefont {Sasaki}},  \emph {et~al.},\ }\href {\doibase
  10.1103/PhysRevD.82.014501} {\bibfield  {journal} {\bibinfo  {journal}
  {Phys.Rev.}\ }\textbf {\bibinfo {volume} {D82}},\ \bibinfo {pages} {014501}
  (\bibinfo {year} {2010})},\ \Eprint {http://arxiv.org/abs/1003.3387}
  {arXiv:1003.3387 [hep-lat]} \BibitemShut {NoStop}%
\bibitem [{\citenamefont {Bratt}\ \emph {et~al.}(2010)\citenamefont {Bratt}
  \emph {et~al.}}]{Bratt:2010jn}%
  \BibitemOpen
  \bibfield  {author} {\bibinfo {author} {\bibfnamefont {J.~D.}\ \bibnamefont
  {Bratt}} \emph {et~al.} (\bibinfo {collaboration} {LHPC Collaboration}),\
  }\href {\doibase 10.1103/PhysRevD.82.094502} {\bibfield  {journal} {\bibinfo
  {journal} {Phys.Rev.}\ }\textbf {\bibinfo {volume} {D82}},\ \bibinfo {pages}
  {094502} (\bibinfo {year} {2010})},\ \Eprint {http://arxiv.org/abs/1001.3620}
  {arXiv:1001.3620 [hep-lat]} \BibitemShut {NoStop}%
\bibitem [{\citenamefont {H\"agler}\ \emph {et~al.}(2008)\citenamefont
  {H\"agler} \emph {et~al.}}]{Hagler:2007xi}%
  \BibitemOpen
  \bibfield  {author} {\bibinfo {author} {\bibfnamefont {P.}~\bibnamefont
  {H\"agler}} \emph {et~al.} (\bibinfo {collaboration} {LHPC}),\ }\href
  {\doibase 10.1103/PhysRevD.77.094502} {\bibfield  {journal} {\bibinfo
  {journal} {Phys. Rev.}\ }\textbf {\bibinfo {volume} {D77}},\ \bibinfo {pages}
  {094502} (\bibinfo {year} {2008})},\ \Eprint {http://arxiv.org/abs/0705.4295}
  {arXiv:0705.4295 [hep-lat]} \BibitemShut {NoStop}%
\bibitem [{\citenamefont {Bernard}\ \emph {et~al.}(2001)\citenamefont
  {Bernard}, \citenamefont {Burch}, \citenamefont {Orginos}, \citenamefont
  {Toussaint}, \citenamefont {DeGrand} \emph {et~al.}}]{Bernard:2001av}%
  \BibitemOpen
  \bibfield  {author} {\bibinfo {author} {\bibfnamefont {C.~W.}\ \bibnamefont
  {Bernard}}, \bibinfo {author} {\bibfnamefont {T.}~\bibnamefont {Burch}},
  \bibinfo {author} {\bibfnamefont {K.}~\bibnamefont {Orginos}}, \bibinfo
  {author} {\bibfnamefont {D.}~\bibnamefont {Toussaint}}, \bibinfo {author}
  {\bibfnamefont {T.~A.}\ \bibnamefont {DeGrand}},  \emph {et~al.},\ }\href
  {\doibase 10.1103/PhysRevD.64.054506} {\bibfield  {journal} {\bibinfo
  {journal} {Phys.Rev.}\ }\textbf {\bibinfo {volume} {D64}},\ \bibinfo {pages}
  {054506} (\bibinfo {year} {2001})},\ \Eprint
  {http://arxiv.org/abs/hep-lat/0104002} {arXiv:hep-lat/0104002 [hep-lat]}
  \BibitemShut {NoStop}%
\bibitem [{\citenamefont {Bistrovi\'c}(2005)}]{Bistrovic_thesis}%
  \BibitemOpen
  \bibfield  {author} {\bibinfo {author} {\bibfnamefont {B.}~\bibnamefont
  {Bistrovi\'c}},\ }\emph {\bibinfo {title} {Perturbative renormalization of
  proton observables in lattice QCD using Domain Wall fermions}},\ \href
  {http://hdl.handle.net/1721.1/32304} {Ph.D. thesis},\ \bibinfo  {school}
  {MIT} (\bibinfo {year} {2005})\BibitemShut {NoStop}%
\bibitem [{\citenamefont {Colangelo}\ \emph {et~al.}(2011)\citenamefont
  {Colangelo}, \citenamefont {D\"urr}, \citenamefont {J\"uttner}, \citenamefont
  {Lellouch}, \citenamefont {Leutwyler} \emph {et~al.}}]{Colangelo:2010et}%
  \BibitemOpen
  \bibfield  {author} {\bibinfo {author} {\bibfnamefont {G.}~\bibnamefont
  {Colangelo}}, \bibinfo {author} {\bibfnamefont {S.}~\bibnamefont {D\"urr}},
  \bibinfo {author} {\bibfnamefont {A.}~\bibnamefont {J\"uttner}}, \bibinfo
  {author} {\bibfnamefont {L.}~\bibnamefont {Lellouch}}, \bibinfo {author}
  {\bibfnamefont {H.}~\bibnamefont {Leutwyler}},  \emph {et~al.},\ }\href
  {\doibase 10.1140/epjc/s10052-011-1695-1} {\bibfield  {journal} {\bibinfo
  {journal} {Eur.Phys.J.}\ }\textbf {\bibinfo {volume} {C71}},\ \bibinfo
  {pages} {1695} (\bibinfo {year} {2011})},\ \Eprint
  {http://arxiv.org/abs/1011.4408} {arXiv:1011.4408 [hep-lat]} \BibitemShut
  {NoStop}%
\bibitem [{\citenamefont {Bazavov}\ \emph {et~al.}(2009)\citenamefont {Bazavov}
  \emph {et~al.}}]{Bazavov:2009fk}%
  \BibitemOpen
  \bibfield  {author} {\bibinfo {author} {\bibfnamefont {A.}~\bibnamefont
  {Bazavov}} \emph {et~al.} (\bibinfo {collaboration} {MILC Collaboration}),\
  }\href@noop {} {\bibfield  {journal} {\bibinfo  {journal} {PoS}\ }\textbf
  {\bibinfo {volume} {CD09}},\ \bibinfo {pages} {007} (\bibinfo {year}
  {2009})},\ \Eprint {http://arxiv.org/abs/0910.2966} {arXiv:0910.2966
  [hep-ph]} \BibitemShut {NoStop}%
\bibitem [{\citenamefont {Bazavov}\ \emph {et~al.}(2010)\citenamefont
  {Bazavov}, \citenamefont {Bernard}, \citenamefont {DeTar}, \citenamefont
  {Du}, \citenamefont {Freeman} \emph {et~al.}}]{Bazavov:2010yq}%
  \BibitemOpen
  \bibfield  {author} {\bibinfo {author} {\bibfnamefont {A.}~\bibnamefont
  {Bazavov}}, \bibinfo {author} {\bibfnamefont {C.}~\bibnamefont {Bernard}},
  \bibinfo {author} {\bibfnamefont {C.}~\bibnamefont {DeTar}}, \bibinfo
  {author} {\bibfnamefont {X.}~\bibnamefont {Du}}, \bibinfo {author}
  {\bibfnamefont {W.}~\bibnamefont {Freeman}},  \emph {et~al.},\ }\href@noop {}
  {\bibfield  {journal} {\bibinfo  {journal} {PoS}\ }\textbf {\bibinfo {volume}
  {LATTICE2010}},\ \bibinfo {pages} {083} (\bibinfo {year} {2010})},\ \Eprint
  {http://arxiv.org/abs/1011.1792} {arXiv:1011.1792 [hep-lat]} \BibitemShut
  {NoStop}%
\bibitem [{\citenamefont {McNeile}\ \emph {et~al.}(2010)\citenamefont
  {McNeile}, \citenamefont {Davies}, \citenamefont {Follana}, \citenamefont
  {Hornbostel},\ and\ \citenamefont {Lepage}}]{McNeile:2010ji}%
  \BibitemOpen
  \bibfield  {author} {\bibinfo {author} {\bibfnamefont {C.}~\bibnamefont
  {McNeile}}, \bibinfo {author} {\bibfnamefont {C.}~\bibnamefont {Davies}},
  \bibinfo {author} {\bibfnamefont {E.}~\bibnamefont {Follana}}, \bibinfo
  {author} {\bibfnamefont {K.}~\bibnamefont {Hornbostel}}, \ and\ \bibinfo
  {author} {\bibfnamefont {G.}~\bibnamefont {Lepage}},\ }\href {\doibase
  10.1103/PhysRevD.82.034512} {\bibfield  {journal} {\bibinfo  {journal}
  {Phys.Rev.}\ }\textbf {\bibinfo {volume} {D82}},\ \bibinfo {pages} {034512}
  (\bibinfo {year} {2010})},\ \Eprint {http://arxiv.org/abs/1004.4285}
  {arXiv:1004.4285 [hep-lat]} \BibitemShut {NoStop}%
\bibitem [{\citenamefont {D\"urr}\ \emph
  {et~al.}(2011{\natexlab{b}})\citenamefont {D\"urr}, \citenamefont {Fodor},
  \citenamefont {Hoelbling}, \citenamefont {Katz}, \citenamefont {Krieg} \emph
  {et~al.}}]{Durr:2010vn}%
  \BibitemOpen
  \bibfield  {author} {\bibinfo {author} {\bibfnamefont {S.}~\bibnamefont
  {D\"urr}}, \bibinfo {author} {\bibfnamefont {Z.}~\bibnamefont {Fodor}},
  \bibinfo {author} {\bibfnamefont {C.}~\bibnamefont {Hoelbling}}, \bibinfo
  {author} {\bibfnamefont {S.}~\bibnamefont {Katz}}, \bibinfo {author}
  {\bibfnamefont {S.}~\bibnamefont {Krieg}},  \emph {et~al.},\ }\href {\doibase
  10.1016/j.physletb.2011.05.053} {\bibfield  {journal} {\bibinfo  {journal}
  {Phys.Lett.}\ }\textbf {\bibinfo {volume} {B701}},\ \bibinfo {pages} {265}
  (\bibinfo {year} {2011}{\natexlab{b}})},\ \Eprint
  {http://arxiv.org/abs/1011.2403} {arXiv:1011.2403 [hep-lat]} \BibitemShut
  {NoStop}%
\bibitem [{\citenamefont {Walker-Loud}\ \emph {et~al.}(2009)\citenamefont
  {Walker-Loud}, \citenamefont {Lin}, \citenamefont {Richards}, \citenamefont
  {Edwards}, \citenamefont {Engelhardt} \emph {et~al.}}]{WalkerLoud:2008bp}%
  \BibitemOpen
  \bibfield  {author} {\bibinfo {author} {\bibfnamefont {A.}~\bibnamefont
  {Walker-Loud}}, \bibinfo {author} {\bibfnamefont {H.-W.}\ \bibnamefont
  {Lin}}, \bibinfo {author} {\bibfnamefont {D.}~\bibnamefont {Richards}},
  \bibinfo {author} {\bibfnamefont {R.}~\bibnamefont {Edwards}}, \bibinfo
  {author} {\bibfnamefont {M.}~\bibnamefont {Engelhardt}},  \emph {et~al.},\
  }\href {\doibase 10.1103/PhysRevD.79.054502} {\bibfield  {journal} {\bibinfo
  {journal} {Phys.Rev.}\ }\textbf {\bibinfo {volume} {D79}},\ \bibinfo {pages}
  {054502} (\bibinfo {year} {2009})},\ \Eprint {http://arxiv.org/abs/0806.4549}
  {arXiv:0806.4549 [hep-lat]} \BibitemShut {NoStop}%
\bibitem [{\citenamefont {Edwards}\ and\ \citenamefont
  {Jo\'o}(2005)}]{Edwards:2004sx}%
  \BibitemOpen
  \bibfield  {author} {\bibinfo {author} {\bibfnamefont {R.~G.}\ \bibnamefont
  {Edwards}}\ and\ \bibinfo {author} {\bibfnamefont {B.}~\bibnamefont {Jo\'o}}
  (\bibinfo {collaboration} {SciDAC, LHPC, and UKQCD Collaborations}),\ }\href
  {\doibase 10.1016/j.nuclphysbps.2004.11.254} {\bibfield  {journal} {\bibinfo
  {journal} {Nucl.Phys.Proc.Suppl.}\ }\textbf {\bibinfo {volume} {140}},\
  \bibinfo {pages} {832} (\bibinfo {year} {2005})},\ \Eprint
  {http://arxiv.org/abs/hep-lat/0409003} {arXiv:hep-lat/0409003 [hep-lat]}
  \BibitemShut {NoStop}%
\bibitem [{\citenamefont {Martinelli}\ \emph {et~al.}(1995)\citenamefont
  {Martinelli}, \citenamefont {Pittori}, \citenamefont {Sachrajda},
  \citenamefont {Testa},\ and\ \citenamefont {Vladikas}}]{Martinelli:1994ty}%
  \BibitemOpen
  \bibfield  {author} {\bibinfo {author} {\bibfnamefont {G.}~\bibnamefont
  {Martinelli}}, \bibinfo {author} {\bibfnamefont {C.}~\bibnamefont {Pittori}},
  \bibinfo {author} {\bibfnamefont {C.~T.}\ \bibnamefont {Sachrajda}}, \bibinfo
  {author} {\bibfnamefont {M.}~\bibnamefont {Testa}}, \ and\ \bibinfo {author}
  {\bibfnamefont {A.}~\bibnamefont {Vladikas}},\ }\href {\doibase
  10.1016/0550-3213(95)00126-D} {\bibfield  {journal} {\bibinfo  {journal}
  {Nucl. Phys.}\ }\textbf {\bibinfo {volume} {B445}},\ \bibinfo {pages} {81}
  (\bibinfo {year} {1995})},\ \Eprint {http://arxiv.org/abs/hep-lat/9411010}
  {arXiv:hep-lat/9411010} \BibitemShut {NoStop}%
\bibitem [{\citenamefont {Gracey}(2003)}]{Gracey:2003yr}%
  \BibitemOpen
  \bibfield  {author} {\bibinfo {author} {\bibfnamefont {J.~A.}\ \bibnamefont
  {Gracey}},\ }\href {\doibase 10.1016/S0550-3213(03)00335-3} {\bibfield
  {journal} {\bibinfo  {journal} {Nucl. Phys.}\ }\textbf {\bibinfo {volume}
  {B662}},\ \bibinfo {pages} {247} (\bibinfo {year} {2003})},\ \Eprint
  {http://arxiv.org/abs/hep-ph/0304113} {arXiv:hep-ph/0304113} \BibitemShut
  {NoStop}%
\bibitem [{\citenamefont {Tiburzi}\ and\ \citenamefont
  {Walker-Loud}(2006)}]{Tiburzi:2005na}%
  \BibitemOpen
  \bibfield  {author} {\bibinfo {author} {\bibfnamefont {B.~C.}\ \bibnamefont
  {Tiburzi}}\ and\ \bibinfo {author} {\bibfnamefont {A.}~\bibnamefont
  {Walker-Loud}},\ }\href {\doibase 10.1016/j.nuclphysa.2005.08.013} {\bibfield
   {journal} {\bibinfo  {journal} {Nucl.Phys.}\ }\textbf {\bibinfo {volume}
  {A764}},\ \bibinfo {pages} {274} (\bibinfo {year} {2006})},\ \Eprint
  {http://arxiv.org/abs/hep-lat/0501018} {arXiv:hep-lat/0501018 [hep-lat]}
  \BibitemShut {NoStop}%
\bibitem [{\citenamefont {Detmold}\ \emph {et~al.}(2002)\citenamefont
  {Detmold}, \citenamefont {Melnitchouk},\ and\ \citenamefont
  {Thomas}}]{Detmold:2002nf}%
  \BibitemOpen
  \bibfield  {author} {\bibinfo {author} {\bibfnamefont {W.}~\bibnamefont
  {Detmold}}, \bibinfo {author} {\bibfnamefont {W.}~\bibnamefont
  {Melnitchouk}}, \ and\ \bibinfo {author} {\bibfnamefont {A.~W.}\ \bibnamefont
  {Thomas}},\ }\href {\doibase 10.1103/PhysRevD.66.054501} {\bibfield
  {journal} {\bibinfo  {journal} {Phys.Rev.}\ }\textbf {\bibinfo {volume}
  {D66}},\ \bibinfo {pages} {054501} (\bibinfo {year} {2002})},\ \Eprint
  {http://arxiv.org/abs/hep-lat/0206001} {arXiv:hep-lat/0206001 [hep-lat]}
  \BibitemShut {NoStop}%
\end{thebibliography}%
